\documentclass[12pt,hidelinks]{article}
\usepackage{mathrsfs,amsmath,epsfig,amssymb,color,graphicx,bm,setspace,setspace}
\usepackage{algorithm,algorithmic,natbib,subcaption,hyperref,lineno}
\usepackage[margin=1in]{geometry}
\usepackage[symbol]{footmisc}
\graphicspath{{figures/}}
\allowdisplaybreaks

\newtheorem{theorem}{Theorem}

\newcommand{\bSigma}{\boldsymbol \Sigma}
\newcommand{\bbeta}{\boldsymbol \beta}
\newcommand{\hbeta}{\hat{\boldsymbol \beta}_{\mbox{\tiny \rm  FULL}}}
\newcommand{\OLSbeta}{\hat{\boldsymbol \beta}_{\mbox{\tiny \rm  OLS}}}
\newcommand{\SCADbeta}{\hat{\boldsymbol \beta}_{\mbox{\tiny \rm  SCAD}}}

\newcommand{\bPsi}{\boldsymbol \Psi}
\newcommand{\bGamma}{\boldsymbol \Gamma}

\newcommand{\X}{\mathbf{X}}

\newcommand{\bUpsilon}{\boldsymbol \Upsilon}

\begin{document}
\title{Refitted cross-validation estimation for high-dimensional subsamples from
  low-dimension full data}
\author{Haixiang Zhang$^{1}$ and  HaiYing Wang$^{2}$\\
  \small \it $^{1}$Center for Applied Mathematics, Tianjin
  University, Tianjin 300072, China\\
  \small \it $^{2}$Department of Statistics, University of Connecticut, Storrs,
  Mansfield, CT 06269, USA }
\maketitle

\setstretch{1.3}

\begin{abstract}
  The technique of subsampling has been extensively employed to address the
  challenges posed by limited computing resources and meet the needs for
  expedite data analysis.  Various subsampling methods have been developed to
  meet the challenges characterized by a large sample size ($n$) with a small
  number of parameters ($p$), by analyzing a subsample of size $r$ such that
  $n\gg r\gg p$.  However, direct applications of these subsampling methods may
  not be suitable when the dimension is also high and available computing
  facilities at hand are only able to analyze a subsample of size similar or
  even smaller than the dimension. In this case, although there is no
  high-dimensional problem in the full data ($n\gg p$), the subsample may have a
  sample size smaller or smaller than the number of parameters, making it a
  high-dimensional problem. We call this scenario the high-dimensional subsample
  from low-dimension full data problem.  In this paper, we tackle this problem
  by proposing a novel subsampling-based approach that combines penalty-based
  dimension reduction and refitted cross-validation. The asymptotic normality of
  the refitted cross-validation subsample estimator is established, which plays
  a crucial role in statistical inference. The proposed method demonstrates
  appealing performance in numerical experiments on simulated data and a real
  data application.

\end{abstract}

\pagebreak
\setstretch{1.8}

\section{Introduction}

Due to advancements in science and technology, the prevalence of big data with massive sample sizes is observed across various fields that require meticulous analysis. The application of traditional statistical methods to such voluminous data presents challenges in terms of both computer memory and computational efficiency. To address these challenges, subsampling has recently gained increasing attention and undergone extensive research. 
The primary concept of a subsampling scheme is to assign appropriate subsampling probabilities to each data point and draw a representative small-sized subsample from the original data. Then, we employ this subsample with appropriate adjustments to ensure some desired properties
for conducting statistical inference.  As a result, subsampling can significantly reduce the computational burden by downsizing the data volume. In recent years, a large number of papers dedicated to the development of innovative subsampling methods tailored for large datasets have emerged. For example, \cite{PingMa2014-JMLR} evaluated the statistical properties of algorithmic leveraging  in the context of selecting representative subsamples. \cite{Wang2018-JASA} and \cite{wang-JLMR-2019} studied the topic on optimal subsampling for logistic regression with massive datasets. \cite{Wang2019-JASA} proposed an information-based optimal subdata selection
for big data linear model. \cite{ZNR-2021-JCGS} considered  an optimal sampling strategy for generalized linear models under measurement constraints. \cite{HTYZ-2020-AOS} presented a  local uncertainty sampling approach for large-scale multiclass logistic regression. \cite{Zhang-Wang2021}, \cite{Poisson-JASA-2021} and \cite{Zuo-2021-CS} studied optimal subsampling for distributed massive datasets. \cite{AFT-sim-2022} proposed an optimal subsampling algorithm to fast approximate the maximum likelihood estimator for parametric accelerate failure time models with massive survival data. \cite{JCGS-Cox} and \cite{rare-Cox-2023} provided optimal subsample selection approaches for large-scale Cox regression. \cite{DCS-2024} proposed the optimal decorrelated score subsampling for generalized linear models with massive data.
The two reviewing papers by \cite{OSP-review-2021} and \cite{SP-review} provide further literature on subsampling.

The computational and storage burden becomes significantly onerous when dealing with datasets characterized by large sample sizes and high-dimensional covariates. The majority of subsampling methods focused on the challenges posed by a large sample size ($n$) relative to a small number of covariates ($p$), by analyzing a subsample of size $r$ such that $n\gg r\gg p$. The direct applications of these subsampling methods may not be suitable when the dimension is high and the available computing facilities can only analyze a subsample of similar or even smaller size than the dimension. The full data does not present a high-dimensional problem in this situation, but the subsample may have a sample size smaller or comparable to the number of parameters, thus posing a challenge in high-dimensional analysis. We refer to this scenario as  the high-dimensional subsample from low-dimension full data issue.
In this paper, we propose
a refitted cross-validation subsample estimation procedure  for large-scale and high-dimensional 
regression, such as  linear, logistic, and Cox models. The key idea of our method is to randomly split the full data into two  halves, do variable selection by using a uniform subsample of the first half of the data set and achieve dimension reduction, and use a subsample of the second half to parallelly fit low dimensional models by appending
 each predictor to the selected set, one at a time, to obtain the estimated coefficient for each
 predictor, regardless of being selected or not, and vice versa. The estimator
proposed is just the average of these two estimators.
 The key benefits of our approach encompass: First, the proposed subsampling method exhibits high computational speed, effectively alleviating the computational bottleneck associated with massive datasets characterized by large $n$ and $p$ simultaneously. Second, the asymptotic normality of subsample-based estimators with non-uniform sampling probabilities is established, thereby addressing a formidable challenge in the context of high-dimensional data. Third, the proposed subsampling framework is universally applicable to a wide range of regression models. i.e., the concept behind our approach is potentially to provide a unified solution for addressing large-scale regression problems in practical applications.

The remainder of this article is organized as follows: In Section \ref{sec-2}, we provide an overview of fundamental notations and definitions pertaining to high-dimensional regression framework, such as linear model, logistic model, and Cox model.
In Section \ref{sec-3}, we present a refitted cross-validation subsample estimation procedure, together with the asymptotic properties of subsample-based estimators. The numerical simulations and real data example are presented in Sections \ref{NS-5} and \ref{sec-6}, respectively. The concluding remarks are presented in Section \ref{sec-7}. All proof details are provided in the Appendix.

\section{Notations and Models} \label{sec-2}

The outcome of interest is denoted as $Y$, and the vector of covariates is
represented by $\mathbf{X} = (X_1,\cdots,X_p)^\prime$.  The existence of a
regression model is assumed, which relates $\mathbf{X}$ to $Y$ and is
characterized by p-dimensional regression coefficients denoted as $\bbeta =
(\beta_1,\cdots,\beta_p)^\prime$. The primary emphasis of our research lies on the simultaneous consideration of massive sample size and high-dimensional covariates. The covariates are assumed to be high-dimensional; however, the dimensionality $p$ is still smaller than the full data sample size $n$. Denote the set of active variables as $\Omega = \{j: \beta_j\neq 0, j=1,\cdots,p\}$. The cardinality of $ \Omega$ is typically small, which satisfies the common sparsity requirement of high-dimensional data. The parameter estimate can be obtained by minimizing a criterion function:
\begin{eqnarray}\label{FULL-1}
\hbeta=\arg\min_{\bbeta} \left\{\frac{1}{n}\sum_{i=1}^n m(\mathbf{Z}_i, \bbeta)\right\},
\end{eqnarray}
where $ \mathbf{Z}_i = (\mathbf{X}_i, Y_i)$ denotes $i$-th observation,
$i=1,\cdots,n$; $\hbeta$ is also referred to as the M-estimator. e.g., the
criterion function is $m(\mathbf{Z}_i, \bbeta) = (Y_i - \mathbf{X}_i^\prime
\bbeta)^2$ for the linear model. 


Note that the SCAD-penalized estimator \cite[]{scad-JASA} is not taken into consideration when deriving the full data estimator given in (\ref{FULL-1}). This decision is based on the following facts: (i) The enhancement in estimation efficiency resulting from the SCAD-penalized technique may not be evident when dealing with a sufficiently large dataset. (ii)  The computational speed of the SCAD-penalized estimator is significantly slower compared to non-penalty based methods, particularly when dealing with massive datasets. (iii)  The SCAD-penalized method only guarantees asymptotic normality for non-zero estimators, while the standard errors of zero-estimators remain unavailable. The unpenalized estimator provides standard errors for all parameters, which are highly valuable for conducting statistical inference.  The following section presents a straightforward simulation to exemplify the aforementioned statements within the framework of a linear model. To be specific, we generate random samples from $Y=\mathbf{X}^\prime \bbeta + \epsilon$, where $\bbeta = (\beta_1,\cdots,\beta_p)^\prime$ is the vector of regression coefficients, and $\epsilon$ is the random error. The components of $\mathbf{X}$ are independent  uniform random variables over $(-1, 1)$, the error term $\epsilon$ follows from $N(0,1)$. We set $\beta_1=\cdots=\beta_{20}=0.5$, and 
$\beta_i=0$ for others; the observations are denoted as $(\mathbf{X}_i, Y_i)$, $i=1,\cdots,n$. The 
ordinary  least square  (OLS) estimator and  SCAD-penalized estimator are 
\begin{eqnarray*}
\OLSbeta=\arg\min_{\bbeta} \left\{\frac{1}{n}\sum_{i=1}^n (Y_i - \mathbf{X}_i^\prime \bbeta)^2\right\},
\end{eqnarray*}
and
\begin{eqnarray*}
\SCADbeta=\arg\min_{\bbeta} \left\{\frac{1}{n}\sum_{i=1}^n (Y_i - \mathbf{X}_i^\prime \bbeta)^2 + \sum_{j=1}^p  P_{\lambda}(\beta_j)\right\},
\end{eqnarray*}
where  $P_{\lambda}(\cdot)$ is the SCAD penalty \cite[]{scad-JASA}. Without loss of generality, we focus on the  squared deviation of interested $\hat{\beta}_1$ in the $l$th repetition of the simulation, which is given as $SD_{l} = (\hat{\beta}^{(l)}_1-\beta_1)^2$. The computational burden necessitates the utilization of only 100 repetitions for calculating the mean of $SD_{l}$, denoted as ${\rm MSD}=\sum_{l=1}^{100}SD_l/100$. The computational speed of $\OLSbeta$ and $\SCADbeta$ is also compared by reporting the average CPU times (in seconds). The dimension of covariates is selected as $p=500$, while the sample size is chosen as $n=550$,  $n=10^5$ and  $n=10^6$, respectively. The results of MSD and CPU in Table \ref{FULL-com-1} are presented, where the estimators $\OLSbeta$ and $\SCADbeta$ are obtained using the popular R functions {\tt lm()} and {\tt ncvreg()}, respectively. The MSD of the SCAD method exhibits a significant improvement compared to that of the OLS method with $n=550$. Under $n=10^5$ and  $n=10^6$, the MSD of the SCAD method is  comparable to that of the OLS method, while the computational efficiency of SCAD is significantly slower than OLS.

\begin{table}[htp] 
\begin{center}
    \caption{The comparison between OLS and SCAD estimators with full data}\label{FULL-com-1}
    \vspace{0.1in} 
    \begin{tabular}{lccccccccccccc}
      \hline
     & \multicolumn{2}{c}{$n=550$} &  & \multicolumn{2}{c}{$n=10^5$} &  & \multicolumn{2}{c}{$n=10^6$}\\
      \cline{2-3}\cline{5-6}\cline{8-9}
     method &MSD &CPU & & MSD &CPU& & MSD &CPU\\
      \hline
     OLS  & 0.05104   &0.073 &&$2.884\times 10^{-5}$  &20.820   &&$3.730\times 10^{-6} $  &241.899\\
     SCAD & 0.02745    &2.645 && $2.833\times 10^{-5} $ &253.384 && $3.729\times 10^{-6} $  &1095.225\\
     \hline
  \end{tabular}
  \end{center}
\end{table}

Estimating the parameters  and quantifying the uncertainty of
 estimators are crucial tasks in practical applications. 
As shown in Table \ref{FULL-com-1}, the huge sample size makes it challenging to calculate full-data estimator due to the limited computing resources. To deal with this issue, subsampling is an effective approach to
reduce the computation and storage burden arising from massive data (\citeauthor{Wang2018-JASA}, \citeyear{Wang2018-JASA}; \citeauthor{IEEE-Poisson}, \citeyear{IEEE-Poisson}; \citeauthor{JCGS-Cox}, \citeyear{JCGS-Cox}).  The direct application of these subsampling methods is not suitable  when the dimension is also high and available computing
  facilities at hand are only able to analyze a subsample of size similar or
  even smaller than the dimension. The reason is that these subsampling methods necessitate a significantly larger subsample size in comparison to the dimension of covariates. However, in the case of a large number of parameters, the subsample may have a smaller sample size, thereby presenting a high-dimensional problem. We refer to this scenario as the issue of obtaining a high-dimensional subsample from low-dimensional full data. In the next section, we propose a refitted cross-validation subsampling  algorithm designed to handle massive datasets with large values of both $n$ and $p$ simultaneously.

\section{Refitted Cross-Validation Subsample Estimators  } \label{sec-3}
The computational and storage burden becomes substantial when performing statistical inference in the face of massive datasets characterized by a large sample size $n$ and high dimensionality $p$.
To address these key concerns, we propose an innovative refitted cross-validation subsample estimation procedure, encompassing the subsequent three steps:

{\bf Step 1}. Randomly partition the full dataset $\mathcal{D}_n = \{\mathbf{Z}_i\}_{i=1}^n$ into two non-overlapping subsets, denoted as $\mathcal{D}_{n_1} = \{\mathbf{Z}_i^{(1)}\}_{i=1}^{n_1}$ and $\mathcal{D}_{n_2} = \{\mathbf{Z}_i^{(2)}\}_{i=1}^{n_2}$, where $|\mathcal{D}^{}_{n_1}|=[n/2]$ and $|\mathcal{D}_{n_2}^{}|=n-n_1$. The $\mathcal{D}_{n_1}$ can be firstly utilized for conducting variable selection in order to identify the active variables. 
However, the computational algorithm for variable selection is time-consuming due to the large sample size of $\mathcal{D}_{n_1}$. To reduce the computational burden when conducting variable selection, we select a uniform subsample  $\mathcal{D}^*_{r_{01}} = \{\mathbf{Z}_i^{0*(1)}\}_{i=1}^{r_{01}}$ without replacement from  $\mathcal{D}^{}_{n_1}$, where  $\mathbf{Z}_i^{0*(1)} = (\mathbf{X}_i^{0*(1)}, Y_i^{0*(1)})$, 
the subsample size  $r_{01}$ is much smaller than $n_1$. 
    Utilizing  the subsample $\mathcal{D}^*_{r_{01}}$, we can derive the SCAD-penalized estimators $\{\hat{\beta}_j^{SCAD}\}_{j=1}^p$ by minimizing this criterion function:
\begin{eqnarray}
L(\bbeta;\mathcal{D}^*_{r_{01}}) = \frac{1}{r_{01}} \sum_{i=1}^{r_{01}}
m(\mathbf{Z}_i^{0*(1)}, \bbeta) + \sum_{j=1}^p  P_{\lambda}(\beta_j),
\end{eqnarray}
where $P_{\lambda}(\cdot)$ is a penalty function (e.g., SCAD penalty; \citeauthor{scad-JASA}, \citeyear{scad-JASA}); the index of nonzero coefficients (active variables) is estimated as 
$\hat{\Omega}^{(1)} = \{j:\hat{\beta}_j^{SCAD}\neq 0, j=1,\cdots,p\}$.  By Theorem 1 of \cite{PM-JASA-2008}, the index set  $\hat{\Omega}_1^{(1)}$ satisfies the selection consistency $\mathbb{P}(\hat{\Omega}^{(1)} = \Omega)\rightarrow 1$, as $r_{01} \rightarrow \infty$.    

{\bf Step 2}. The sub-model is refitted based on $\mathcal{D}_{n_2}$, using the selected variables from $\hat{\Omega}^{(1)}$.
Then we can obtain the corresponding  criterion function
\begin{eqnarray}\label{rfitLM}
M^{(2)}(\bbeta_{\hat{\Omega}^{(1)}}) = \frac{1}{n_2}\sum_{i=1}^{n_2} m\left(\mathbf{Z}_{i,\hat{\Omega}^{(1)}}^{(2)}, \bbeta_{\hat{\Omega}^{(1)}}\right),
\end{eqnarray}
where  the sub-vector $\bbeta_{\hat{\Omega}^{(1)}}$ represents a subset of $\bbeta$ consisting of indices belonging to $\hat{\Omega}^{(1)}$; $\mathbf{Z}_{i,\hat{\Omega}^{(1)}}^{(2)} = (\mathbf{X}_{i,\hat{\Omega}^{(1)}}^{(2)}, Y_i^{(2)})$. The minimization of (\ref{rfitLM}) constitutes a low-dimensional optimization problem; however, the sample size  $n_2$ remains substantial. The computational burden remains heavy when directly handling $\mathcal{D}_{n_2}$. To expedite the calculation efficiency, we further randomly take a subsample 
${\mathcal{D}}_{2,r}^{*} = \{\mathbf{Z}_{i,\hat{\Omega}^{(1)}}^{*(2)}
\}_{i=1}^{r_{}}$ from $\mathcal{D}_{n_2}$ with replacement, where
$\mathbf{Z}_{i,\hat{\Omega}^{(1)}}^{*(2)} =
(\mathbf{X}_{i,\hat{\Omega}^{(1)}}^{*(2)}, Y_i^{*(2)})$. Following
\cite{IEEE-Poisson},  the practical optimal sampling probabilities are 
\begin{eqnarray}
\pi_i^{(2)} = (1-\delta)\frac{\left\|\dot{m}\left(\mathbf{Z}_{i,\hat{\Omega}^{(1)}}^{(2)}, \check{\bbeta}_{\hat{\Omega}^{(1)}}\right)\right\|}{\sum_{j=1}^{n_2}\left\|\dot{m}\left(\mathbf{Z}_{j,\hat{\Omega}^{(1)}}^{(2)}, \check{\bbeta}_{\hat{\Omega}^{(1)}}\right)\right\|} + \delta\frac{1}{n_2}, ~i=1,\cdots,n_2,
\end{eqnarray}
where $\delta=0.1$, and $\check{\bbeta}_{\hat{\Omega}^{(1)}}$ is a pilot
estimator from (\ref{rfitLM}) using a uniform subsample of size $r_{02}$. For
example with linear regression, the optimal sampling probabilities have the expressions:
\begin{eqnarray*}
\pi_i^{(2)} = (1-\delta)\frac{|Y_i^{(2)} - \check{\bbeta}_{\hat{\Omega}^{(1)}}^\prime\mathbf{X}_{i,\hat{\Omega}^{(1)}}^{(2)}|\|\mathbf{X}_{i,\hat{\Omega}^{(1)}}^{(2)}\|}
{\sum_{j=1}^{n_2}|Y_j^{(2)} - \check{\bbeta}_{\hat{\Omega}^{(1)}}^\prime\mathbf{X}_{j,\hat{\Omega}^{(1)}}^{(2)}|\|\mathbf{X}_{j,\hat{\Omega}^{(1)}}^{(2)}\|} + \delta\frac{1}{n_2}, ~i=1,\cdots,n_2.
\end{eqnarray*}

 In order to provide consistent estimators for all $p$ parameters $\beta_i$'s, we adopt a selection-assisted partial regression approach (\citeauthor{FZBL-2019}, \citeyear{FZBL-2019}; \citeauthor{FL-JMLR-2021}, \citeyear{FL-JMLR-2021}; \citeauthor{HDECR-2023}, \citeyear{HDECR-2023}).  The partial regression can be conducted by employing $ \mathbf{X}_{\hat{\Omega}_{+j}^{(1)}}^{}$ as the predictor for $Y$,
where  $\hat{\Omega}_{+j}^{(1)} = \hat{\Omega}^{(1)}\cup\{j\}$ for $j=1,\cdots,p$. Based on ${\mathcal{D}}_{2,r}^{*}$ and resultant subsampling probabilities $\{\pi_i^{*(2)}\}_{i=1}^r$, we can derive a subsample estimator as follows:
\begin{eqnarray}
\hat{\bbeta}_{\hat{\Omega}_{+j}^{(1)}} = \arg\min_{\bbeta_{\hat{\Omega}_{+j}^{(1)}}} \left\{\frac{1}{r}\sum_{i=1}^r \frac{1}{\pi_i^{*(2)}}m\left( \mathbf{Z}_{i,\hat{\Omega}_{+j}^{(1)}}^{*(2)}, \bbeta_{\hat{\Omega}_{+j}^{(1)}} \right)\right\}.
\end{eqnarray}
Note that the estimator $\hat{\bbeta}_{\hat{\Omega}_{+j}^{(1)}}$ is based on a subsample (${\mathcal{D}}_{2,r}^{*}$) extracted from $\mathcal{D}_{n_2}$, whereas the corresponding least squares estimator used by \cite{FL-JMLR-2021} relies on the complete dataset $\mathcal{D}_{n_2}$.

Based on \cite{IEEE-Poisson},  we denote the asymptotic sampling ratio  as $\rho = 2\lim_{n\rightarrow \infty}r/n$, the asymptotic normality of $\hat{\bbeta}_{\hat{\Omega}_{+j}^{(1)}}$ is
\begin{eqnarray*}
\hat{\bSigma}^{-1/2}\{\hat{\bbeta}_{\hat{\Omega}_{+j}^{(1)}} - {\bbeta}_{\hat{\Omega}_{+j}^{(1)}}\}\stackrel{d}{\longrightarrow} N(0,\mathbf{I}),
\end{eqnarray*}
 where $\stackrel{d}{\longrightarrow}$ denotes convergence in distribution, 
$\hat{\bSigma} = {\hat{\bGamma}}^{-1}\hat{\bPsi} \hat{\bGamma}^{-1}$ with
\begin{eqnarray*}
\hat{\bGamma}&=&\frac{1}{n_2r}\sum_{i=1}^r\frac{1}{\pi_i^{*(2)}}
\ddot{m}\left( \mathbf{Z}_{i,\hat{\Omega}_{+j}^{(1)}}^{*(2)}, \hat{\bbeta}_{\hat{\Omega}_{+j}^{(1)}} \right)\\
\hat{\bPsi} &=&  \frac{1}{n^2_2r^2}\sum_{i=1}^r\frac{1}{\{\pi_i^{*(2)}\}^2}\dot{m}\left( \mathbf{Z}_{i,\hat{\Omega}_{+j}^{(1)}}^{*(2)}, \hat{\bbeta}_{\hat{\Omega}_{+j}^{(1)}} \right)\dot{m}\left( \mathbf{Z}_{i,\hat{\Omega}_{+j}^{(1)}}^{*(2)}, \hat{\bbeta}_{\hat{\Omega}_{+j}^{(1)}} \right)^\prime\\
&& - \rho \frac{1}{n_2r^2}\sum_{i=1}^r\frac{1}{\pi_i^{*(2)}}\dot{m}\left( \mathbf{Z}_{i,\hat{\Omega}_{+j}^{(1)}}^{*(2)}, \hat{\bbeta}_{\hat{\Omega}_{+j}^{(1)}} \right)\dot{m}\left( \mathbf{Z}_{i,\hat{\Omega}_{+j}^{(1)}}^{*(2)}, \hat{\bbeta}_{\hat{\Omega}_{+j}^{(1)}} \right)^\prime.
\end{eqnarray*}
We denote $\hat{\beta}_{2,j} = \left(\hat{\bbeta}_{\hat{\Omega}_{+j}^{(1)}}\right)_j$ as  the element of $\hat{\bbeta}_{\hat{\Omega}_{+j}^{(1)}}$ corresponding to covariate $X_j$; and $\hat{\sigma}_{2,j}^2 = \hat{\bSigma}_{j,j}$ as the $j$th diagonal element of $\hat{\bSigma}$.


{\bf Step 3}. Similarly, the dimensionality of covariates is reduced by performing variable selection with SCAD using a uniform subsample of $\mathcal{D}^{}_{n_2}$, resulting in the set of selected active variables $\hat{\Omega}^{(2)}$. Refit a sub-model using the selected variables in $\hat{\Omega}^{(2)}$. The subsample estimator $\hat{\beta}_{1,j}$ and its corresponding standard error $\hat{\sigma}_{1,j}$ are calculated based on a subsample of $\mathcal{D}^{}_{n_1}$, similar to Step 2. The resulting subsample-based estimator is constructed as
\begin{eqnarray}\label{PE-3}
\hat{\beta}_j = \frac{\hat{\beta}_{1,j} + \hat{\beta}_{2,j}}{2},~~j=1,\cdots,p.
\end{eqnarray}

\begin{figure}[!tpb]
\centerline{\includegraphics[scale = 0.45]{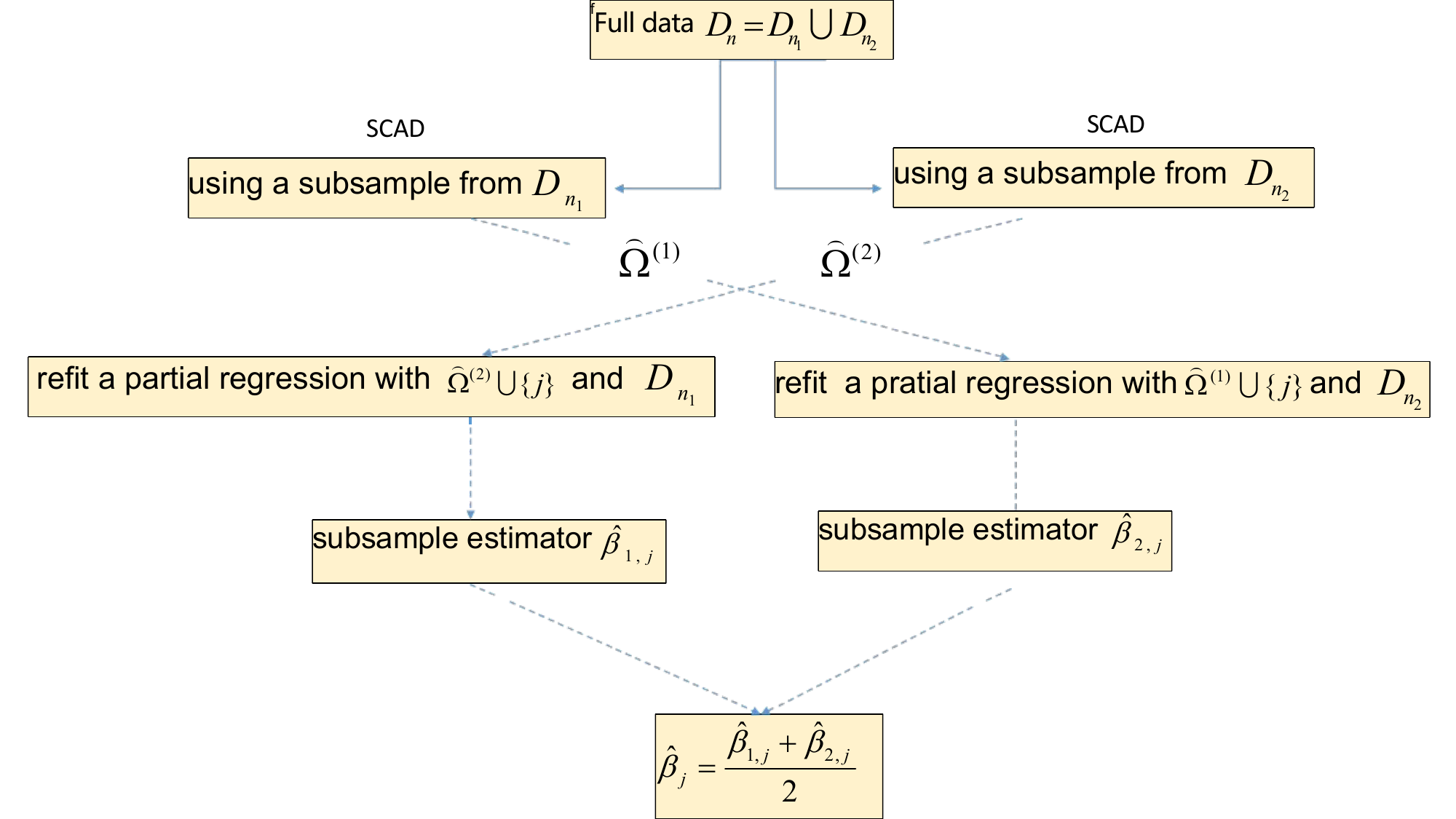}}
\vspace{0.3cm}
\begin{center}
 \caption{ A scenario of refitted cross-validation subsampling-based estimation procedure.}
 \label{fig:1}
\end{center}
\end{figure}

In Figure \ref{fig:1}, we present a schematic diagram for the refitted cross-validation subsample estimation procedure. The computational burden of parameter estimation has been significantly alleviated by employing subsampling, which will be evaluated in the simulation section. The asymptotic properties and variance estimation, in addition to the point estimates provided in (\ref{PE-3}), are crucial for statistical inference. The subsequent theorem will explore the asymptotic normality of subsampling-based estimators $\hat{\beta}_j$'s. 

\begin{theorem}\label{th1}
 Under the assumptions 1-4 of \cite{IEEE-Poisson}, as $r\rightarrow \infty$ and $n\rightarrow \infty$, then  we have
 \begin{eqnarray}\label{AN-6}
\hat{\sigma}_j^{-1}(\hat{\beta}_j - \beta_j) \stackrel{d}{\longrightarrow} N(0,1),
 \end{eqnarray}
 where $\hat{\sigma}_j^2 = (\hat{\sigma}_{1,j}^2 + \hat{\sigma}_{2,j}^2)/4$ is an estimated variance of $\hat{\beta}_j$ for $j=1,\cdots,p$. 
\end{theorem}

The asymptotic normality established in equation (\ref{AN-6}) guarantees that
the estimator $\hat{\beta}_j$ is consistent for the true parameter $\beta_j$ as
the sample size $r \rightarrow \infty$. In the context of statistical inference,
constructing confidence intervals is of great interest. Leveraging Theorem
\ref{th1}, we can construct a 95\% confidence interval for $\beta_j$ ($j=1,\cdots,p$) as follows:
\begin{eqnarray*}
CI_j=[\hat{\beta}_j-1.96\hat{\sigma}_j,~\hat{\beta}_j+1.96\hat{\sigma}_j].
\end{eqnarray*}

\section{Numerical Simulation}
\label{NS-5}
In this section, we conduct some simulations to evaluate the performance of our proposed method. We generate random samples from the following three models, (i) linear model: $Y=\mathbf{X}^\prime \bbeta + \epsilon$, where the true parameter is $\bbeta= (1,0.8,  0.75, -0.5,  0.75,0,\cdots,0)^\prime$, the error term follows from $N(0,1)$; (ii) logistic model: $P(Y=1) = {\exp(\mathbf{X}^\prime \bbeta)}/\{1+\exp(\mathbf{X}^\prime \bbeta)\}$, where  the true parameter is $\bbeta= (0.75, 1.25, 1.5, 0.85, 1.2, 0,\cdots,0)^\prime$; (iii) Cox model: 
$\lambda(t|\mathbf{X}) = \lambda_0(t) \exp(\mathbf{X}^\prime \bbeta)$, where the true parameter is $\bbeta= (1,1.5,0.85,0.75,2, 0,\cdots,0)^\prime$, the baseline hazard
function is $\lambda_0(t)=0.5t$. The censoring times $C_i$'s are independently generated from a uniform distribution over $(0, c_0)$ with $c_0$ being chosen so that the censoring rate (CR) is about 30\%. We consider two cases for the generation of covariate $\mathbf{X}=(X_1,\cdots,X_p)^\prime$ with dimension $p=300$:\\
 {\it Case} 1 : components of $\mathbf{X}$ are independent  uniform random variables over $(-1, 1)$.\\
 {\it Case} 2: $\mathbf{X}$ follows $0.5N(-\mathbf{1},\bUpsilon)+0.5N(\mathbf{1},\bUpsilon)$, where $\Upsilon_{jk}=0.5^{|j-k|}$, i.e., $\X$ follows a mixture of two multivariate normal distributions.

 For comparison, we also consider using uniform sampling in Step 2 of our method (denoted as ``UNIF"), while our proposed method is denoted as ``OSP". All the results are based on 500 repetitions, where the full data size is  $n=10^6$,  the subsample size is chosen as 
In Tables 1-6, we report the subsample estimation results for $\{\beta_i\}_{i=1}^6$, including 
bias (BIAS) given by the mean of point estimates minus the true value, the sample standard deviation (SSD) of point estimates, the average of estimated standard errors (ESE), the empirical coverage of probability (CP) of 95\% confidence interval; Other $\beta_i$'s have similar performances with that of $\beta_6$, and thus not included.  The unbiasedness of both UNIF and OSP estimators can be observed from Tables 1-6, with the SSD and ESE exhibiting close proximity. The empirical coverage probabilities are approximately 0.95, indicating reasonable asymptotic normality of the subsample estimator in practical applications. 

\begin{figure}[htp] 
  \centering
  \begin{subfigure}{0.39\textwidth}
    \includegraphics[width=\textwidth]{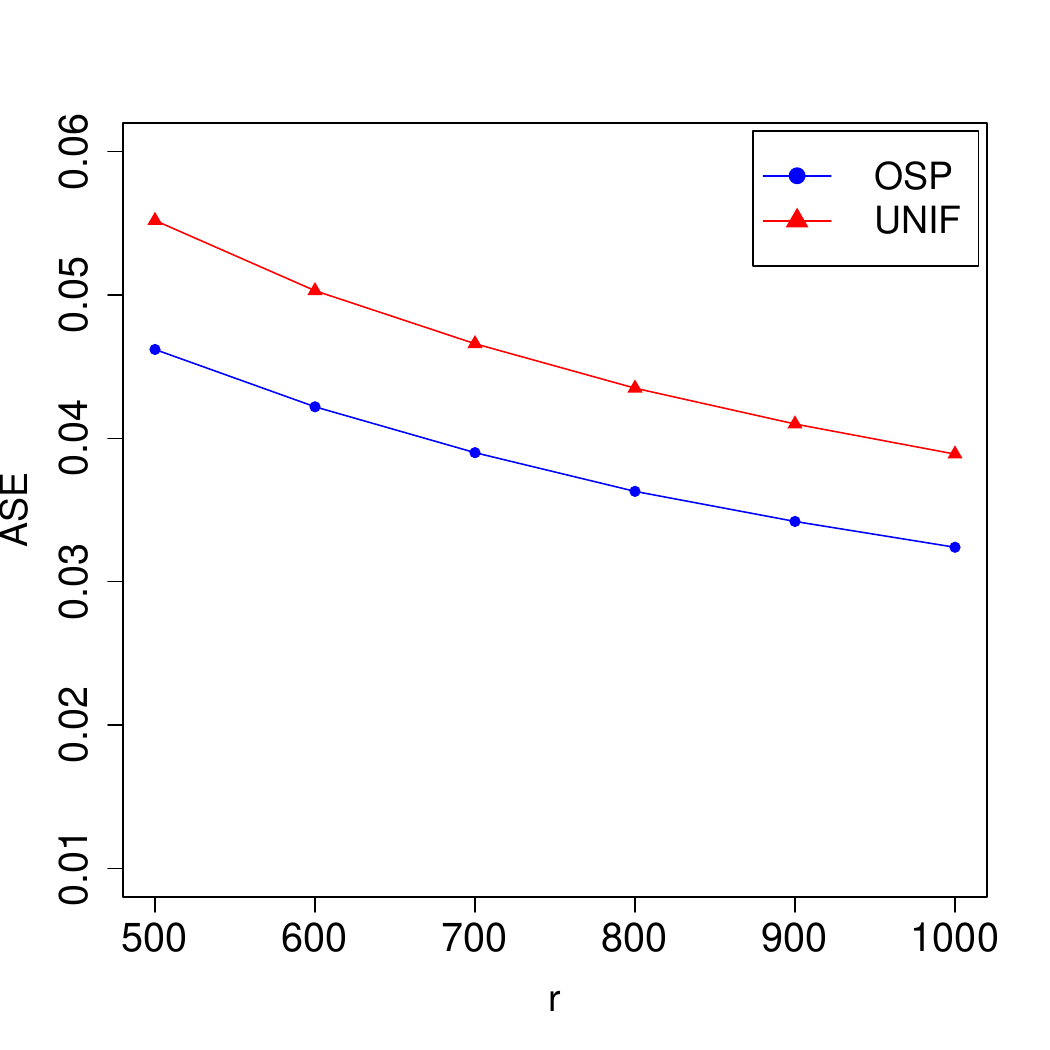}
    \caption{linear model with case 1.}
  \end{subfigure}
  \begin{subfigure}{0.39\textwidth}
    \includegraphics[width=\textwidth]{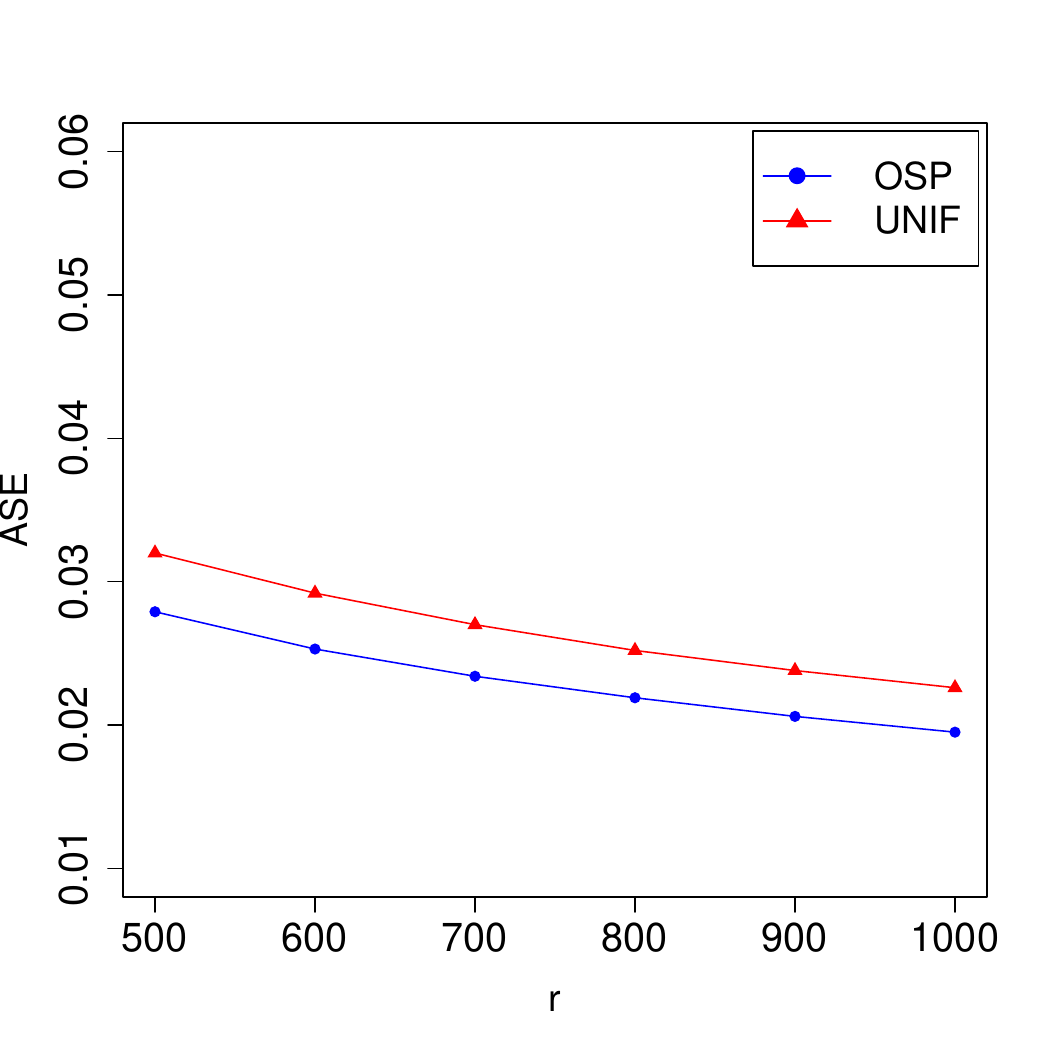}
    \caption{linear model with case 2.}
  \end{subfigure}
    \begin{subfigure}{0.39\textwidth}
    \includegraphics[width=\textwidth]{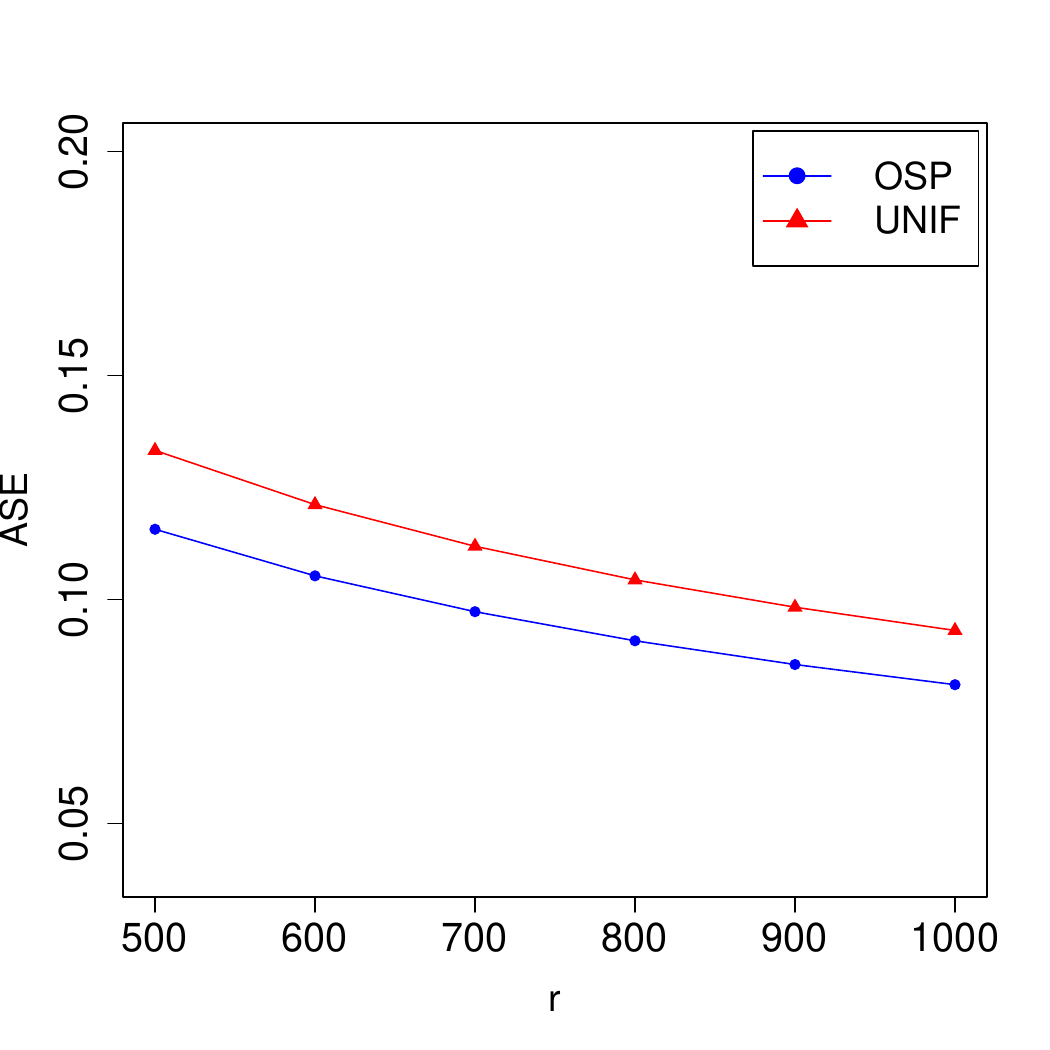}
    \caption{logistic model with case 1.}
  \end{subfigure}
  \begin{subfigure}{0.39\textwidth}
    \includegraphics[width=\textwidth]{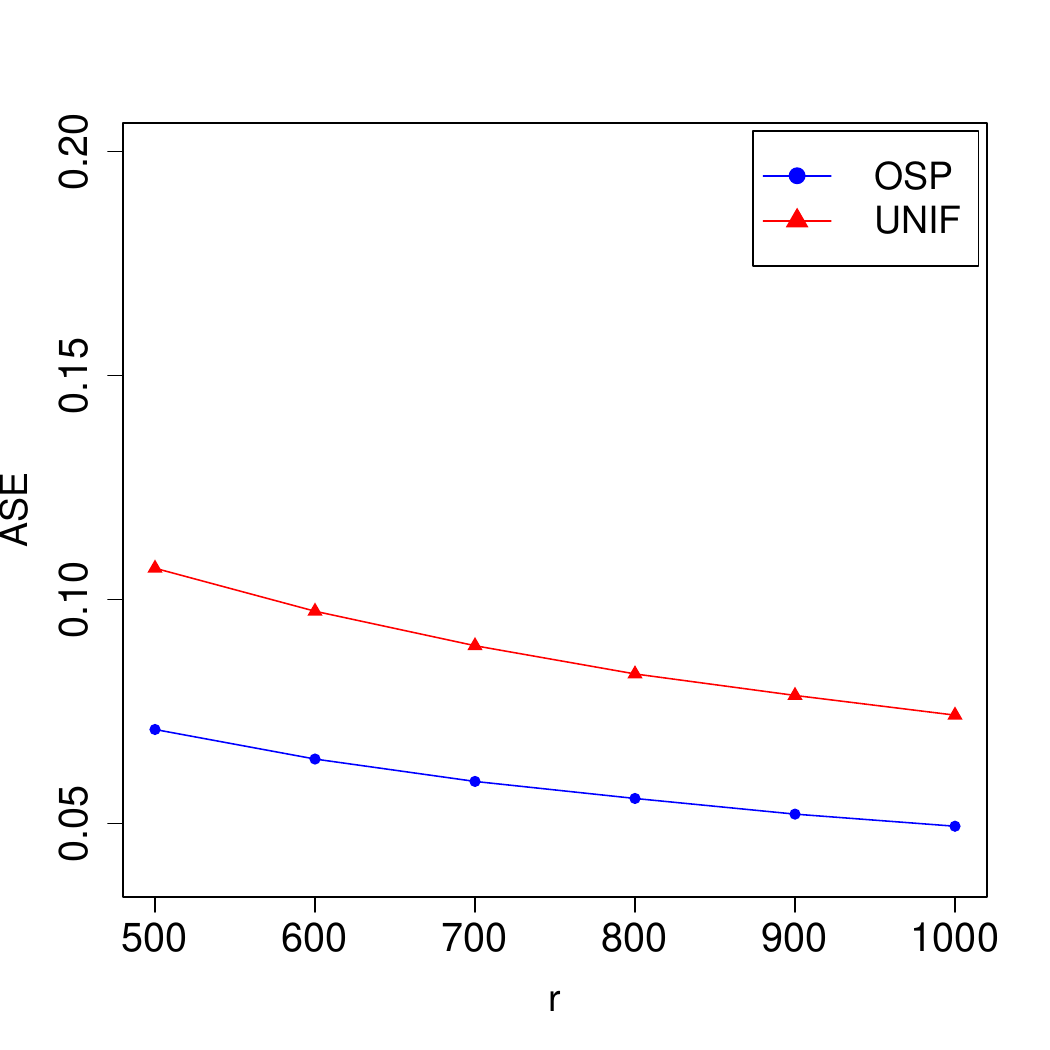}
    \caption{logistic model with case 2.}
  \end{subfigure}
    \begin{subfigure}{0.39\textwidth}
    \includegraphics[width=\textwidth]{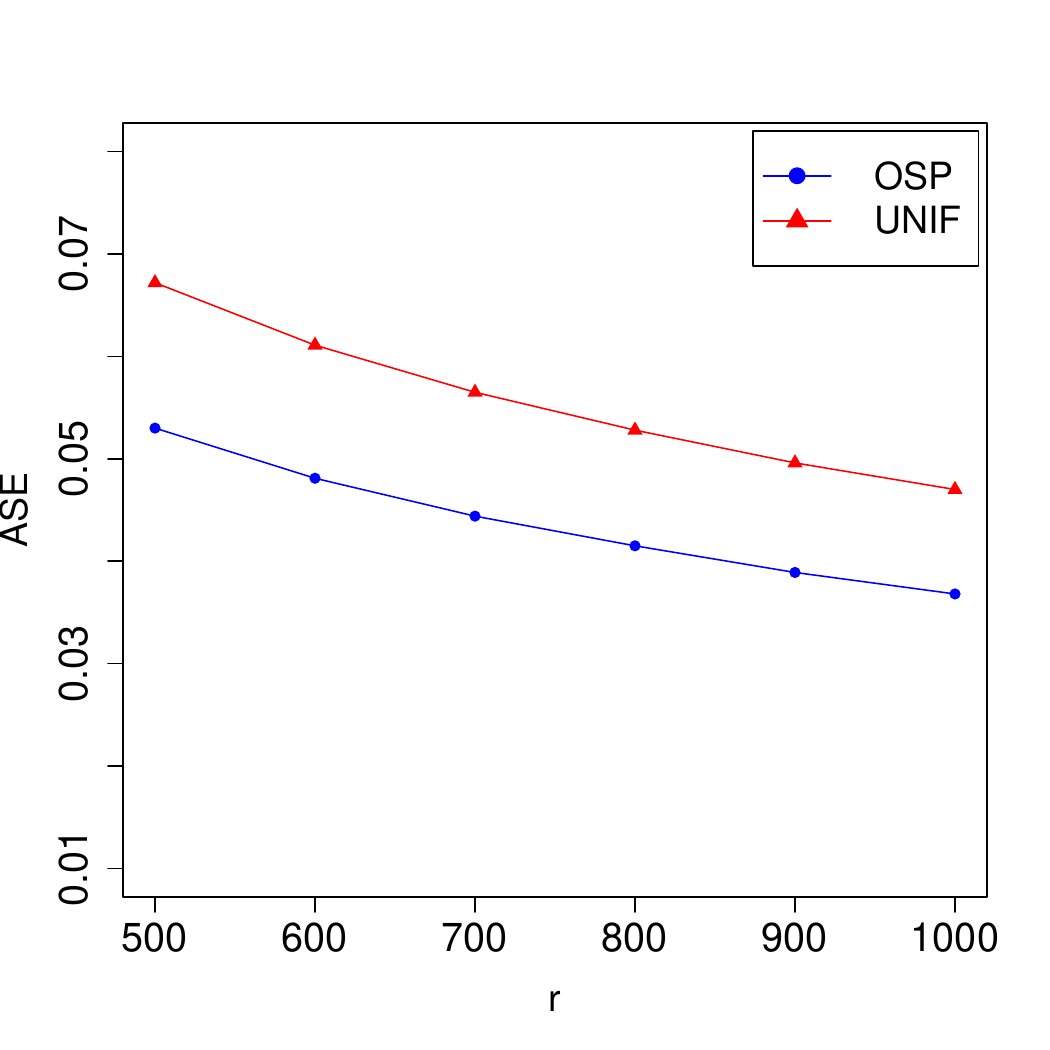}
    \caption{Cox model with case 1.}
  \end{subfigure}
  \begin{subfigure}{0.39\textwidth}
    \includegraphics[width=\textwidth]{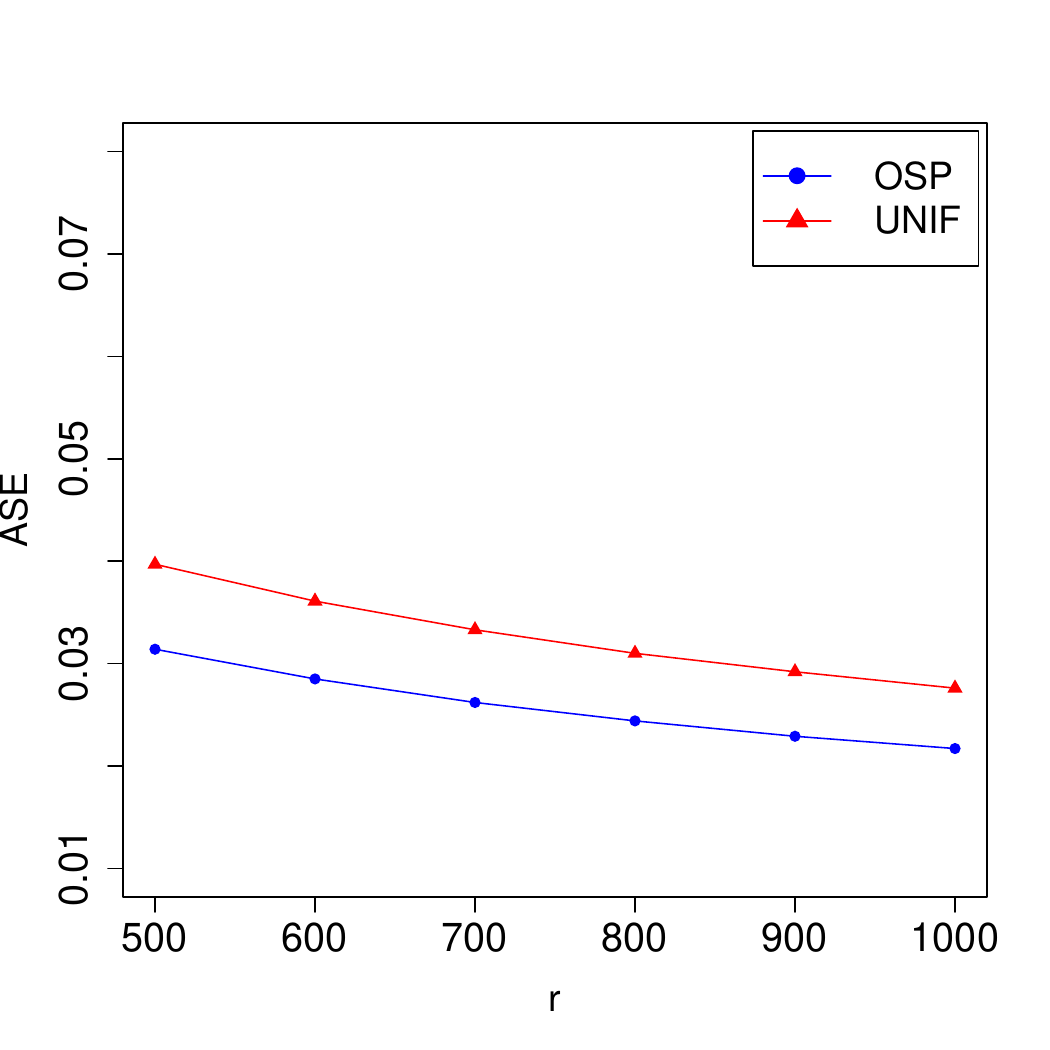}
    \caption{Cox model with case 2.}
  \end{subfigure}
 \vspace{-0.1cm}
\begin{center}
  \caption{ The ASEs for different subsampling methods.}
  \label{fig:2}
\end{center}
\end{figure}

\begin{table}[htp] 
  \begin{center}
    \caption{Simulation results of the subsample estimators with linear model (Case 1).}
    \label{tab:L1}
    \vspace{0.1in} \small
    \begin{tabular}{lccccccccccc}
      \hline
      & &  & \multicolumn{4}{c}{OSP} &  & \multicolumn{4}{c}{UNIF} \\
      \cline{4-7}\cline{9-12}
      & $\beta_j$ & &Bias &SSD &ESE &CP & & Bias &SSD &ESE &CP\\
      \hline
      $r=500$
     & $\beta_1$ & &-0.0043  & 0.0483 &0.0441 &0.930 &&0.0009 & 0.0558 & 0.0551 &0.952 \\
      & $\beta_2$ & &0.0050  & 0.0462 &0.0443 &0.946 &&0.0009 & 0.0525 & 0.0551 &0.962 \\
      & $\beta_3$ & &-0.0001 & 0.0450 &0.0443 &0.950 &&0.0005 & 0.0532 & 0.0551 &0.956 \\
      & $\beta_4$ & &-0.0021 & 0.0427 &0.0441 &0.956 &&-0.0014 &0.0563 & 0.0551 &0.948 \\
      & $\beta_5$ & &0.0014  & 0.0464 &0.0443 &0.934 &&-0.0026 &0.0540 &0.0552 &0.958\\
      & $\beta_6$ & &-0.0003 & 0.0455 &0.0462 &0.956 &&-0.0039 &0.0550 &0.0551 &0.944 \\
          \hline
      $r=800$
  & $\beta_1$     & &-0.0050 & 0.0335 &0.0347 &0.954 && -0.0039 & 0.0449 &0.0435 & 0.950 \\
      & $\beta_2$ & & 0.0021 & 0.0348 &0.0348 &0.938 && 0.0041  &  0.0431 &0.0435 & 0.952 \\
      & $\beta_3$ & &-0.0012 & 0.0358 &0.0347  &0.940 &&-0.0022 & 0.0448 &0.0434 &0.954 \\
      & $\beta_4$ & &-0.0003 & 0.0367 & 0.0347  &0.936 &&-0.0011 & 0.0420 &0.0434 & 0.958 \\
      & $\beta_5$ & &-0.0006 & 0.0351 & 0.0347  &0.938 &&-0.0025 & 0.0445 &0.0435 &0.956 \\
      & $\beta_6$ & & -0.0016& 0.0364 & 0.0363 &0.952 &&0.0001   & 0.0453 &0.0435 & 0.940 \\
            \hline
      $r=1000$
    & $\beta_1$ & &0.0006    & 0.0314 &0.0310 & 0.952 && 0.0004 & 0.0412 & 0.0389 & 0.926 \\
      & $\beta_2$ & &0.0020  & 0.0312 &0.0311 &0.942  &&0.0034 & 0.0373 & 0.0389 &0.960\\
      & $\beta_3$ & &-0.0007 & 0.0321 &0.0310 &0.946 &&0.0007 & 0.0397 & 0.0388 & 0.948 \\
      & $\beta_4$ & & 0.0022 & 0.0310 &0.0310 &0.952  && 0.0016 &  0.0399 & 0.0388 & 0.936 \\
      & $\beta_5$ & &-0.0016 & 0.0310 &0.0310 &0.958  &&-0.0004 & 0.0386& 0.0389 &0.948 \\
      & $\beta_6$ & &-0.0041 & 0.0336 & 0.0324 & 0.958 && -0.0039 &  0.0364 &0.0388 &0.966 \\
      \hline
    \end{tabular}
  \end{center}
\end{table}

\begin{table}[htp] 
  \begin{center}
    \caption{Simulation results of the subsample estimators with linear model (Case 2).}
    \label{tab:L2}
    \vspace{0.1in} \small
    \begin{tabular}{lccccccccccc}
      \hline
      & &  & \multicolumn{4}{c}{OSP} &  & \multicolumn{4}{c}{UNIF} \\
      \cline{4-7}\cline{9-12}
      & $\beta_j$ & &Bias &SSD &ESE &CP & & Bias &SSD &ESE &CP\\
      \hline
      $r=500$
      & $\beta_1$ & &-0.0007 & 0.0303 &0.0295 &0.944 &&0.0007 & 0.0374 &0.0366  &0.952  \\
      & $\beta_2$ & &0.0012  & 0.0346 &0.0334 &0.952 &&0.0005 & 0.0421 &0.0410  &0.938 \\
      & $\beta_3$ & &-0.0007 & 0.0326 &0.0334 &0.954 &&0.0019 &  0.0398 &0.0410 &0.956 \\
      & $\beta_4$ & &-0.0030 & 0.0321 &0.0335 &0.962 &&-0.0011 & 0.0405 &0.0411 &0.948 \\
      & $\beta_5$ & &0.0005 & 0.0295  &0.0297 &0.960 &&0.0001 &  0.0351 &0.0367 &0.962 \\
      & $\beta_6$ & &-0.0001 &  0.0335&0.0321 &0.946 &&-0.0028 & 0.0365 &0.0367 &0.958 \\
          \hline
      $r=800$
      & $\beta_1$ & &0.0007 & 0.0233 &0.0232 &0.954 && 0.0001 & 0.0304 &0.0289  &0.942 \\
      & $\beta_2$ & &-0.0013 & 0.0260 &0.0264 & 0.960 &&-0.0007 &0.0326 &0.0325 &0.954 \\
      & $\beta_3$ & &0.0001 & 0.0242 &0.0263 &0.964 &&0.0026 & 0.0335 &0.0324  &0.938 \\
      & $\beta_4$ & &-0.0017 &0.0262 &0.0263 &0.960 &&-0.0032 &0.0337 & 0.0324 &0.950 \\
      & $\beta_5$ & &0.0008 & 0.0231 &0.0232 &0.954 &&0.0022 & 0.0286 & 0.0290  &0.948 \\
      & $\beta_6$ & &0.0001 & 0.0239 &0.0251 &0.966 &&-0.0007 &0.0302 & 0.0290  &0.944 \\
            \hline
      $r=1000$
      & $\beta_1$ & &-0.0006 &  0.0208 &0.0207 &0.944 &&-0.0016 & 0.0254 &0.0259 &0.952 \\
      & $\beta_2$ & &-0.0007 &  0.0239 &0.0235 &0.954 &&0.0006 & 0.0296 &0.0289 &0.946 \\
      & $\beta_3$ & &0.0022  & 0.0239 &0.0233 &0.938 && 0.0008 & 0.0292 & 0.0289 &0.948 \\
      & $\beta_4$ & &-0.0033 & 0.0224 &0.0235 &0.950 &&-0.0030 & 0.0285 &0.0290 &0.954 \\
      & $\beta_5$ & & 0.0022 & 0.0193 &0.0207 &0.968 && 0.0018 & 0.0246 &0.0259 & 0.960 \\
      & $\beta_6$ & &-0.0004 & 0.0219 &0.0225 & 0.954 &&-0.0032 & 0.0257&0.0259 &0.958\\
      \hline
    \end{tabular}
  \end{center}
\end{table}


\begin{table}[htp] 
  \begin{center}
    \caption{Simulation results of the subsample estimators with logistic model (Case 1)}
    \label{tab:log1}
    \vspace{0.1in} \small
    \begin{tabular}{lccccccccccc}
      \hline
      & &  & \multicolumn{4}{c}{OSP} &  & \multicolumn{4}{c}{UNIF} \\
      \cline{4-7}\cline{9-12}
      & $\beta_j$ & &Bias &SSD &ESE &CP & & Bias &SSD &ESE &CP\\
      \hline
      $r=500$
      & $\beta_1$ & &-0.0067 & 0.1074 &0.1138 &0.964 &&0.0123 &  0.1374 & 0.1364 &0.944 \\
      & $\beta_2$ & &0.0012 & 0.1173  &0.1189 & 0.952 &&0.0268 & 0.1409 &0.1422 &0.936 \\
      & $\beta_3$ & &-0.0019 & 0.1201 &0.1218 & 0.956 &&0.0300 & 0.1487 & 0.1457  & 0.946 \\
      & $\beta_4$ & &-0.0034 & 0.1210 &0.1146 & 0.934 && 0.0172 &0.1338 &0.1372 &0.954  \\
      & $\beta_5$ & &-0.0093 & 0.1159 &0.1182 & 0.960 &&0.0239 &0.1403 & 0.1415 &0.950 \\
      & $\beta_6$ & &-0.0047 & 0.1132 &0.1154 &0.954 &&-0.0001 & 0.1305 &0.1332 &0.958 \\
          \hline
      $r=800$
   & $\beta_1$    & &-0.0014 & 0.0924 &0.0895 &0.940 &&0.0019 & 0.1099 & 0.1067 &0.950 \\
      & $\beta_2$ & &-0.0047 & 0.0940 &0.0935 &0.958 &&0.0134 & 0.1171 &0.1114 &0.946 \\
      & $\beta_3$ & &-0.0029 & 0.0982 &0.0958 &0.936 &&0.0153 & 0.1092 & 0.1140 & 0.954 \\
      & $\beta_4$ & &0.0044 &  0.0871 &0.0899 &0.960 &&0.0056 & 0.1005 &0.1074 &0.964 \\
      & $\beta_5$ & &-0.0055 & 0.0958 &0.0929 &0.944 &&0.0121 & 0.1141 &0.1108 & 0.954 \\
      & $\beta_6$ & &-0.0088 & 0.0873 &0.0908 &0.958 &&-0.0018 &  0.1067 &0.1043 & 0.936 \\
            \hline
      $r=1000$
     & $\beta_1$  & &0.0001 & 0.0811 & 0.0799 &0.952 &&0.0016 & 0.0986 &0.0953 &0.954 \\
      & $\beta_2$ & &-0.0047 & 0.0837 &0.0831 &0.942 &&0.0011 & 0.0984 &0.0991 &0.946 \\
      & $\beta_3$ & &-0.0075 &  0.0850 & 0.0854 &0.958 &&0.0065 & 0.1032 &0.1016 &0.942 \\
      & $\beta_4$ & &0.0077 &  0.0729 &0.0804 &0.972 &&0.0002 & 0.0951 & 0.0958 & 0.948 \\
      & $\beta_5$ & &-0.0057 & 0.0792 &0.0831 &0.956 &&0.0100 &  0.1031 &0.0989 & 0.942 \\
      & $\beta_6$ & &-0.0022 &  0.0824 & 0.0810 &0.942 &&-0.0125  & 0.0902 &0.0930 & 0.962 \\
      \hline
    \end{tabular}
  \end{center}
\end{table}

\begin{table}[htp] 
  \begin{center}
    \caption{Simulation results of the subsample estimators with logistic model (Case 2)}
    \label{tab:log2}
    \vspace{0.1in} \small
    \begin{tabular}{lccccccccccc}
      \hline
      & &  & \multicolumn{4}{c}{OSP} &  & \multicolumn{4}{c}{UNIF} \\
      \cline{4-7}\cline{9-12}
      & $\beta_j$ & &Bias &SSD &ESE &CP & & Bias &SSD &ESE &CP\\
      \hline
      $r=500$
      & $\beta_1$ & &-0.0039 & 0.0813 &0.0798 & 0.954 && 0.0059 & 0.1317 &0.1298 &0.942 \\
      & $\beta_2$ & &-0.0109 & 0.1427 &0.0981 &0.930 &&0.0451  & 0.1655 &0.1573 &0.934 \\
      & $\beta_3$ & &-0.0066 & 0.1041 & 0.1029 &0.956 &&0.0383 &0.1711 &0.1652 &0.948 \\
      & $\beta_4$ & &0.0046 & 0.0970  &0.0907 &0.956 && 0.0368 & 0.1514  &0.1466 &0.942 \\
      & $\beta_5$ & &0.0033 &  0.0948 &0.0884 &0.966 &&0.0314 & 0.1509 &0.1434 &0.940 \\
      & $\beta_6$ & &0.0065 & 0.0868  &0.0813 &0.934 && -0.0032 & 0.1201 & 0.1224 &0.952 \\
          \hline
      $r=800$
   & $\beta_1$    & &-0.0095 & 0.0606 &0.0622 &0.950 &&0.0118  &0.0998  &0.1019 &0.938 \\
      & $\beta_2$ & &0.0032 & 0.0936 &0.0763 &0.940 &&0.0353   & 0.1215 &0.1229 &0.954 \\
      & $\beta_3$ & &-0.0045 &  0.0815 &0.0807 &0.946 &&0.0251 & 0.1302 &0.1291 &0.960 \\
      & $\beta_4$ & &-0.0019 &  0.0728 &0.0707 &0.944 &&0.0211 & 0.1182 &0.1144 &0.954 \\
      & $\beta_5$ & &-0.0003 & 0.0731 &0.0687 & 0.952 &&0.0248 & 0.1169 &0.1116 &0.936 \\
      & $\beta_6$ & &0.0014 & 0.0614 &0.0636 &0.944 &&-0.0054  & 0.1021 &0.0955 &0.942 \\
            \hline
      $r=1000$
     & $\beta_1$  & &0.0023 & 0.0547 &0.0555 &0.956  &&0.0069 & 0.0926 &0.0905  &0.952 \\
      & $\beta_2$ & &-0.0062 &0.0711 &0.0677 &0.942 &&0.0173 & 0.1108 &0.1090 &0.936 \\
      & $\beta_3$ & &0.0001 & 0.0707 &0.0715 &0.950 &&0.0131 &  0.1229 & 0.1142 &0.924 \\
      & $\beta_4$ & & 0.0081 &0.0623 &0.0628 &0.952 && 0.0213 & 0.1049 &0.1015 & 0.938 \\
      & $\beta_5$ & &0.0001 & 0.0615 &0.0613 &0.948  && 0.0090 & 0.0963 &0.0991 &0.954 \\
      & $\beta_6$ & & -0.0003 & 0.0551&0.0567&0.952  && 0.0015 & 0.0850 &0.0851 &0.950 \\
      \hline
    \end{tabular}
  \end{center}
\end{table}

\begin{table}[htp] 
  \begin{center}
    \caption{Simulation results of the subsample estimators with Cox model (Case 1)}
    \label{tab:cox1}
    \vspace{0.1in} \small
    \begin{tabular}{lccccccccccc}
      \hline
      & &  & \multicolumn{4}{c}{OSP} &  & \multicolumn{4}{c}{UNIF} \\
      \cline{4-7}\cline{9-12}
      & $\beta_j$ & &Bias &SSD &ESE &CP & & Bias &SSD &ESE &CP\\
      \hline
      $r=500$
      & $\beta_1$ & &-0.0010 & 0.0566 &0.0561 & 0.952 &&0.0081 & 0.0730 &0.0728 &0.958 \\
      & $\beta_2$ & &-0.0035 & 0.0605 &0.0629 &0.956 &&0.0118 &  0.0835 &0.0793 & 0.934 \\
      & $\beta_3$ & &0.0047 & 0.0527 &0.0544 &0.964 &&0.0159 & 0.0785 & 0.0710 & 0.926 \\
      & $\beta_4$ & &0.0035 & 0.0519 &0.0534 &0.958 &&0.0063 & 0.0720 &0.0702 &0.952 \\
      & $\beta_5$ & &-0.0052 &0.0667 & 0.0717 &0.962 &&0.0176 & 0.0850 &0.0879 &0.954 \\
      & $\beta_6$ & &-0.0007&0.0531 & 0.0529 &0.938 &&-0.0010 & 0.0665 &0.0671 &0.950 \\
          \hline
      $r=800$
   & $\beta_1$    & &-0.0037 &0.0440 &0.0439 &0.964 &&0.0059 & 0.0577 &0.0571 &0.940 \\
      & $\beta_2$ & &-0.0015 &0.0491 &0.0494 &0.948 &&0.0050 & 0.0625 & 0.0622 &0.956 \\
      & $\beta_3$ & &0.0030 &0.0416 &0.0427 &0.958  &&0.0116 & 0.0540 &0.0558 &0.952 \\
      & $\beta_4$ & &-0.0028 &0.0417 & 0.0419 &0.954  &&0.0002 & 0.0590 & 0.0550 & 0.936 \\
      & $\beta_5$ & &-0.0043 & 0.0597 & 0.0562 &0.920  &&0.0065 & 0.0684 &0.0690 & 0.948 \\
      & $\beta_6$ & &-0.0008 &0.0414 &0.0414 &0.948 && 0.0015 & 0.0522 &0.0526 &0.960 \\
            \hline
      $r=1000$
     & $\beta_1$  & &-0.0029 & 0.0391 &0.0390 &0.954 &&0.0026 & 0.0492 &0.0509 &0.960 \\
      & $\beta_2$ & &-0.0083 & 0.0472 &0.0439 &0.930 &&0.0029 & 0.0546 &0.0555 &0.950 \\
      & $\beta_3$ & &0.0013 & 0.0403 &0.0378  &0.946 &&0.0078 & 0.0505 &0.0497 &0.946 \\
      & $\beta_4$ & &-0.0032 & 0.0359 &0.0372 &0.968 && 0.0005 &0.0514 &0.0491 &0.936 \\
      & $\beta_5$ & &-0.0077 & 0.0522 &0.0499 & 0.948 && 0.0037 & 0.0626 &0.0616 &0.948 \\
      & $\beta_6$ & & -0.0035 & 0.0354 &0.0367 &0.956 &&-0.0035 & 0.0470 &0.0470 & 0.946 \\
      \hline
    \end{tabular}
  \end{center}
\end{table}

\begin{table}[htp] 
  \begin{center}
    \caption{Simulation results of the subsample estimators with Cox model (Case 2)}
    \label{tab:cox2}
    \vspace{0.1in} \small
    \begin{tabular}{lccccccccccc}
      \hline
      & &  & \multicolumn{4}{c}{OSP} &  & \multicolumn{4}{c}{UNIF} \\
      \cline{4-7}\cline{9-12}
      & $\beta_j$ & &Bias &SSD &ESE &CP & & Bias &SSD &ESE &CP\\
      \hline
      $r=500$
      & $\beta_1$ & &-0.0044 & 0.0422 & 0.0414 &0.940 &&0.0079 & 0.0511 &0.0538 &0.962 \\
      & $\beta_2$ & &-0.0054 & 0.0521 &0.0535 &0.962 &&0.0114 & 0.0654 &0.0668 &0.956 \\
      & $\beta_3$ & &-0.0047 & 0.0414 &0.0426 &0.950 && 0.0049 &0.0569  &0.0564 &0.942 \\
      & $\beta_4$ & &-0.0035 & 0.0404 &0.0410 &0.946 &&0.0067 & 0.0525 & 0.0552 & 0.960 \\
      & $\beta_5$ & &-0.0107& 0.0615 &0.0617 &0.934 &&0.0168 &  0.0759 &0.0737 &0.942 \\
      & $\beta_6$ & &0.0019 & 0.0350 &0.0360 &0.956 && 0.0017 & 0.0450 &0.0455 &0.954 \\
          \hline
      $r=800$
   & $\beta_1$    & &-0.0045 & 0.0305 &0.0323 &0.952 &&0.0076 & 0.0419 &0.0421 &0.952 \\
      & $\beta_2$ & &-0.0006 & 0.0405 &0.0419 &0.952 &&0.0071 & 0.0509 &0.0522 &0.952 \\
      & $\beta_3$ & &-0.0039 & 0.0306 &0.0331 &0.960 &&0.0057 &  0.0450&0.0441 & 0.946 \\
      & $\beta_4$ & &-0.0010 & 0.0334 &0.0321 &0.942 &&0.0021 & 0.0453 &0.0431 & 0.946 \\
      & $\beta_5$ & &-0.0048 & 0.0445 &0.0481 &0.964 &&0.0107 & 0.0557 &0.0574 &0.956 \\
      & $\beta_6$ & &-0.0018 & 0.0260 &0.0279 &0.952 &&-0.0023 &0.0341 & 0.0356 &0.956 \\
            \hline
      $r=1000$
     & $\beta_1$  & &-0.0027 & 0.0275 &0.0286 &0.950 &&0.0037 & 0.0376 & 0.0374 &0.946 \\
      & $\beta_2$ & &-0.0032 & 0.0373 &0.0371 &0.950 &&0.0089 &  0.0484 &0.0465 &0.944 \\
      & $\beta_3$ & &-0.0013 & 0.0277 &0.0294 &0.962 &&0.0047 &0.0409 &0.0393   &0.948 \\
      & $\beta_4$ & &-0.0053 & 0.0308 &0.0284 & 0.928 &&0.0015 & 0.0362 &0.0383 &0.962 \\
      & $\beta_5$ & &-0.0061 & 0.0412 &0.0428 & 0.956 &&0.0084 &  0.0515 &0.0512 &0.950 \\
      & $\beta_6$ & &-0.0037 & 0.0245 &0.0248 & 0.956 &&-0.0022 &0.0305 &0.0316 &0.954 \\
      \hline
    \end{tabular}
  \end{center}
\end{table}
The estimated standard error for ${\hat{\beta}}_j$ in the $l$th repetition of the simulation is denoted as $SE_{lj}$, and we define $SE_l = \frac{1}{p}\sum_{j=1}^p SE_{lj}$. The average of $SE_l$ is calculated by performing 500 repetitions of the simulation, denoted as ${\rm ASE}=\sum_{l=1}^{500}SE_l/500$. The ASEs of OSP and UNIF estimators are presented in Figure \ref{fig:2}, where the OSP estimator exhibits a significantly smaller ASE compared to that of the UNIF estimator. Hence, the optimal subsampling method we propose exhibits superior statistical efficiency in terms of ASEs. 

Finally, we conducte a simulation to compare the computational efficiency of UNIF and Lopt methods, with sample sizes of $5\times 10^5$ and $10^6$, and dimensions of 300 and 500, respectively. The full data method has also been regarded as a benchmark, which is computed using the R functions {\tt lm()}, {\tt glm()} and {\tt coxph()} for the linear, logistic and Cox model respectively. In Table \ref{tab:time}, we report the  algorithm's CPU time for Case \uppercase \expandafter {\romannumeral 1} with  $r=500$ (in seconds), where the results are based on the mean time of 10 repetitions. The results demonstrate that the subsampling-based method exhibits significantly higher computational efficiency compared to the full data method. 
The computational advantage of UNIF over Lopt lies in the fact that UNIF does not require the calculation of sampling probabilities. However, the difference in computational times between UNIF and Lopt is not substantial, as the primary computational time of the algorithm lies in the series of refitted partial regressions rather than the calculation of sampling probabilities.

\begin{table}[htp] 
  \begin{center}
    \caption{The CPU time for Case \uppercase \expandafter {\romannumeral 1} with  $r=500$ (in seconds)}
    \label{tab:time}
    \vspace{0.1in} \small
    \begin{tabular}{lllcccccccccc}
      \hline
      & &  & \multicolumn{3}{c}{$p$=300} &  & \multicolumn{3}{c}{$p$=500} \\
      \cline{4-6}\cline{8-10}
    Model  & Sample Size & &UNIF &Lopt &Full Data & & UNIF &Lopt &Full Data\\
      \hline
 Linear& $n=5\times 10^5$ & &1.347 & 1.478 &43.11  &&2.819 &2.978 &114.92  \\
        & $n=10^6$         & &2.343 & 2.402 &94.16  &&4.509 & 4.625 &251.26  \\
          \hline
 Logistic& $n=5\times 10^5$ & &3.157 & 3.245 &187.38  &&17.977 & 18.423 &495.03  \\
        & $n=10^6$         & &4.286 & 4.660 &421.50   &&20.202 & 20.574 &1130.93  \\
          \hline
 Cox &  $n=5\times 10^5$   & &30.382 & 32.324 &302.00  &&64.529 & 70.375 &839.83  \\
        & $n=10^6$         & &31.965 & 33.729 &620.31  &&66.588 & 73.636 &1799.61  \\
          \hline
    \end{tabular}
  \end{center}
   {\vspace{0cm} \hspace{-0.3cm}\footnotesize $\ddag$ ``Full Data":  calculated with  R functions {\tt lm()}, {\tt glm()}  and {\tt coxph()}  for the linear, logistic and Cox model, respectively. }
\end{table}

\section{Real Data Analysis}\label{sec-6}

The field of wave energy is rapidly advancing and holds great promise as a renewable energy source to effectively address the challenges posed by global warming and climate change.  
We conducted an extensive analysis of a vast dataset comprising 49 wave energy converters, utilizing wave scenarios from Perth and Sydney. The dataset is publicly accessible at https://archive.ics.uci.edu/dataset/882/large-scale+wave+energy+farm.  The response variable $Y_i$ represents the total power output of the wave farm, while $\mathbf{X}_i= (X_{i1},\cdots,X_{i98})^\prime$ is a vector of covariates that captures measurements from energy converters with $p=98$. The sample size for this study is $n=54,007$. 

\begin{table}[htp] 
  \begin{center}
    \caption{Estimates, standard errors, and 95\% confidence intervals for $\{\beta_j\}_{j=1}^{5}$ with one subsample.}
    \label{tab:L2}
    \vspace{0.1in} \small
    \begin{tabular}{lccccccccccc}
      \hline
      & &  & \multicolumn{3}{c}{OSP} &  & \multicolumn{3}{c}{UNIF} \\
      \cline{4-6}\cline{8-10}
     & $\beta_j$ & &Est &SE  &CI & & Est &SE  &CI\\
      \hline
      r=500 
      & $\beta_1$ & &0.1547 & 0.0336  &[0.0888, 0.2206]   &&0.0919 & 0.0337  &[0.0259, 0.1581] \\
      & $\beta_2$ & &-0.0294 &0.0189  &[-0.0665,  0.0077] &&0.0322  & 0.0257 &[-0.0182, 0.0825] \\
      & $\beta_3$ & & 0.0750 &  0.0300 &[0.0162, 0.1339]   && 0.1024  & 0.0337 &[0.0364, 0.1684] \\
      & $\beta_4$ & &-0.1076 & 0.0384  &[-0.1829, -0.0322] && -0.0065  & 0.0224 &[-0.0505, 0.0374] \\
      & $\beta_5$ & & -0.0709 & 0.0362  &[-0.1420, 0.0002] && -0.0692 &0.0319 &[-0.1317, -0.0068] \\
      \hline
      r=800 
      & $\beta_1$ & &0.2195 & 0.0235 &[0.1734, 0.2656] &&0.1978  & 0.0261 &[0.1466, 0.2489] \\
      & $\beta_2$ & &-0.0401  &0.0152 &[-0.0698, -0.0103] &&-0.0027  & 0.0216 &[-0.0451, 0.0396] \\
      & $\beta_3$ & &0.1043  &0.0254 &[0.0544, 0.1540] &&0.0905  &0.0259 &[0.0397, 0.1413] \\
      & $\beta_4$ & &-0.0342  &0.0174 &[-0.0684, -$5\times 10^{-5}$] &&0.0015  &0.0181 &[-0.0340, 0.0371] \\
      & $\beta_5$ & &-0.0625  & 0.0274 &[-0.1162, -0.0088] &&-0.0194  &0.0239 &[-0.0662, 0.0274] \\
      \hline
      r=1000 
      & $\beta_1$ & &0.2385 & 0.0213 &[0.1965, 0.2804]    &&0.2196  & 0.0242 &[0.1721, 0.2671] \\
      & $\beta_2$ & &0.0156  &0.0149 &[-0.0135, 0.0447]   &&0.0479  & 0.0184 &[0.0119, 0.0840] \\
      & $\beta_3$ & &0.0275  &0.0295 &[-0.0303, 0.0853]   &&-0.0203  &0.0299 &[-0.0789,  0.0383] \\
      & $\beta_4$ & &-0.0054  &0.0085 &[-0.0221, 0.0114]  &&0.0071  &0.0174 &[-0.0269, 0.0411] \\
      & $\beta_5$ & &-0.0621  &0.0212 &[-0.1037, -0.0206] &&-0.1092 &0.0252 &[-0.1586, -0.0598] \\
      \hline
    \end{tabular}
  \end{center}
\end{table}

\begin{figure}[!tpb]
\centerline{\includegraphics[scale = 0.5]{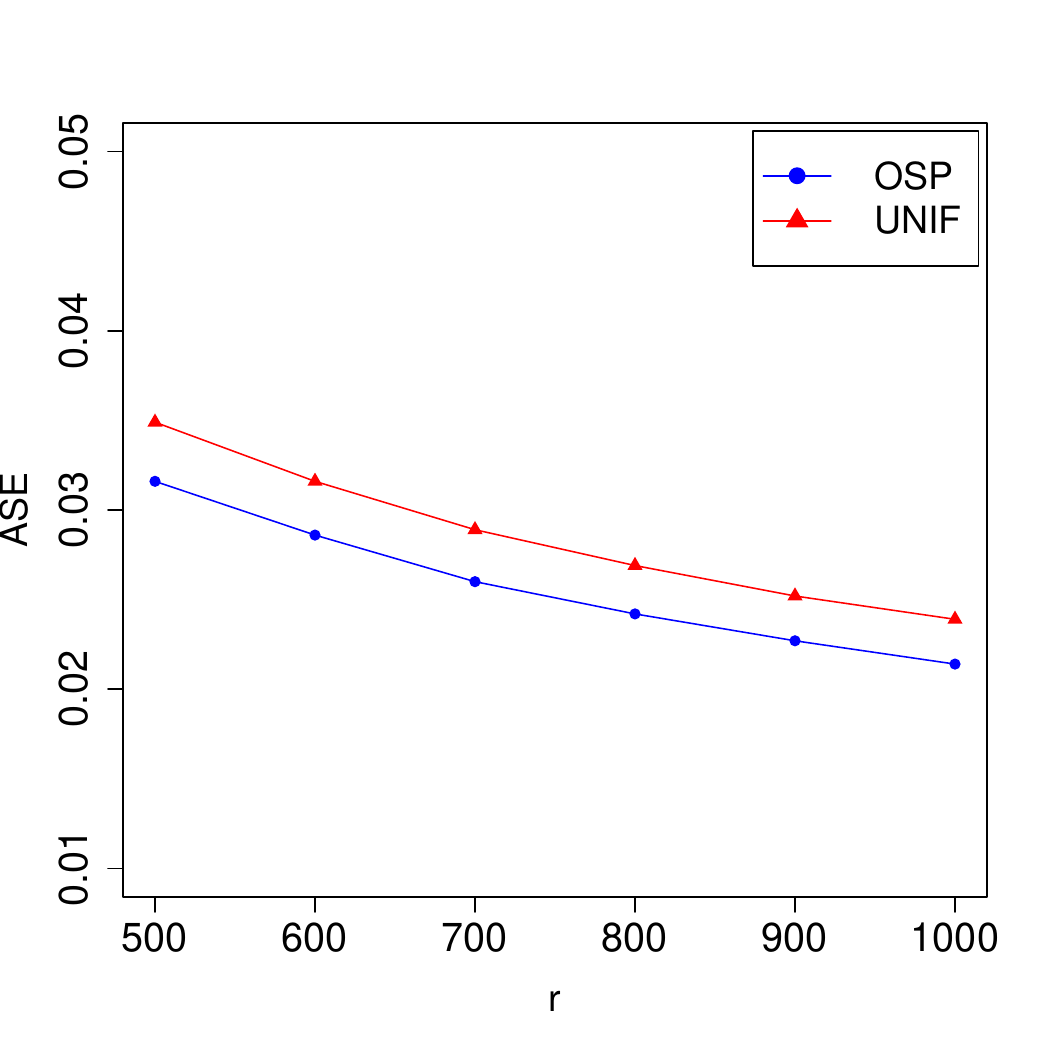}}
\vspace{0.3cm}
\begin{center}
 \caption{ The ASEs for OSP and UNIF methods in real data example.}
 \label{fig:3}
\end{center}
\end{figure}

This dataset is modeled using the linear regression model $Y=\mathbf{X}^\prime \bbeta + \epsilon$, while the OSP and UNIF subsampling methods are employed for parameter estimation of the model, where the subsample size is chosen as $r$=500, 600, 700, 800, 900 and 1000, respectively. In Figure \ref{fig:3}, we present the plot of ASEs for OSP and UNIF, where ASE is defined as in section \ref{NS-5}. The statistical efficiency of the OSP method is significantly superior to that of UNIF, as demonstrated in Figure \ref{fig:3}. Moreover, the estimates, standard errors, and 95\% confidence intervals for $\{\beta_j\}_{j=1}^{5}$ are presented in Table \ref{tab:L2} with one subsample, with respective values of $r=$500, 800, and 1000. For the sake of brevity, the corresponding results for $\{\beta_j\}_{j=6}^{98}$ are omitted in this context.

\section{Concluding Remarks}\label{sec-7}
In this paper, we have proposed a fast subsampling strategy for dealing with massive datasets with 
large $n$ and $p$.  The refitted cross-validation subsample estimators have been derived for large-scale and high-dimensional regression models. The establishment of asymptotic normality was highly advantageous for conducting statistical inference. The utility of our method was demonstrated through simulations and a real data example. The focus of our simulation has primarily been on linear, logistic, and Cox models; however, the proposed framework offers valuable insights for addressing other large-scale and high-dimensional regression models, including quantile regression \cite[]{Wang-MA2020}, additive hazards model \cite[]{zuo2021-sim}, multiclass logistic regression \cite[]{HTYZ-2020-AOS}, longitudinal data regression \cite[]{SLD-2023}, and accelerated failure time model \cite[]{AFT-SC-2024}. The exploration of these subjects warrants additional investigation in the course of our study.

\section*{Appendix}

The proofs of asymptotic normality for subsample-based estimator are provided in the Appendix.


\noindent {\bf Proof of Theorem \ref{th1}}. By \cite{IEEE-Poisson}, the asymptotic normality of $\hat{\bbeta}_{\hat{\Omega}_{+j}^{(1)}}$ is stated as 
\begin{eqnarray*}
\hat{\bSigma}^{-1/2}\{\hat{\bbeta}_{\hat{\Omega}_{+j}^{(1)}} - {\bbeta}_{\hat{\Omega}_{+j}^{(1)}}\}\stackrel{d}{\longrightarrow} N(0,\mathbf{I}).
\end{eqnarray*}
Therefore, it is straightforward to derive that $\hat{\sigma}_{2,j}^{-1}(\hat{\beta}_{2,j} - \beta_j)\stackrel{d}{\rightarrow}N(0,1)$ as $r\rightarrow \infty$ and $n\rightarrow \infty$. The asymptotic normality of $\hat{\beta}_{1,j}$ is obtained in a similar manner, where $\hat{\sigma}_{1,j}^{-1}(\hat{\beta}_{1,j} - \beta_j)$ converges in distribution to $N(0,1)$.
The refitted cross-validation subsample estimation procedure suggests that $\hat{\beta}_{1,j}$ and $\hat{\beta}_{2,j}$ are two estimators that become asymptotically independent. i.e., the two terms hold asymptotically:
\begin{eqnarray*}
E\left(\frac{\hat{\beta}_{1,j} + \hat{\beta}_{2,j}}{2}~\vline~\mathcal{D}_n\right)= \hat{\beta}_{j},
\end{eqnarray*}
and
\begin{eqnarray*}
Var\left(\frac{\hat{\beta}_{1,j} + \hat{\beta}_{2,j}}{2}~\vline~\mathcal{D}_n\right)= \frac{\hat{\sigma}_{1,j}^2 + \hat{\sigma}_{2,j}^2}{4}.
\end{eqnarray*}
Then,  as $r\rightarrow \infty$ and $n\rightarrow \infty$ we have
 \begin{eqnarray*}
\hat{\sigma}_j^{-1}(\hat{\beta}_j - \beta_j) \stackrel{d}{\longrightarrow} N(0,1),
 \end{eqnarray*}
 where $\hat{\sigma}_j^2 = (\hat{\sigma}_{1,j}^2 + \hat{\sigma}_{2,j}^2)/4$, and $\hat{\beta}_j = \{\hat{\beta}_{1,j} + \hat{\beta}_{2,j}\}/{2}$, $j=1,\cdots,p$. This ends the proof.


\bibliographystyle{natbib}
\bibliography{reference}


\end{document}